\documentclass[
aps,%
12pt,%
final,%
notitlepage,%
oneside,%
onecolumn,%
nobibnotes,%
nofootinbib,%
superscriptaddress,%
noshowpacs,%
centertags]%
{revtex4}

\begin{document}

\pagenumbering{arabic}

\title{NEW SETS OF EXPERIMENTAL WAVENUMBER VALUES FOR TRIPLET-TRIPLET ROVIBRONIC TRANSITIONS OF $H_2$ AND $D_2$}

\author{B.~P.~Lavrov}
\email{lavrov@pobox.spbu.ru}
\author{A.~S.~Mikhailov}
\author{I.~S.~Umrikhin}
\affiliation{
Faculty of Physics, St.-Petersburg State University, \\
St.-Petersburg, 198904, Russia}

\begin{abstract}

New sets of experimental wavenumber values for triplet-triplet 
rovibronic transitions of $H_2$ and $D_2$ for visible part of 
spectrum ($400 \div 700$ nm) have been obtained. The digital intensity registration 
providing a linear response of the detector gave us the 
opportunity of digital deconvolution of the recorded line profiles.
For line centers that made it possible to reach an accuracy ($< 0.0006$ nm) limited
only by selfconsistency of various wavenumber standards.
New sets of wavenumber values were obtained with accuracy $0.006 \div 0.05$ cm$^{-1}$.

\end{abstract}

\maketitle

At present almost all available experimental data on rovibronic line 
wavenumbers of $H_2$ and $D_2$ molecules were obtained from emission 
spectra by photographic recording. Our recent studies revealed that 
these data have significant differences with values predicted by 
Rydberg-Ritz combination principle as well as with fragmentary 
experimental data obtained by laser induced fluorescence and with 
our own emission spectroscopy data obtained with photoelectric 
recording by matrix detectors \cite{ALMU2008, LU2008}. 
The minority of the differences is caused by misprints and erroneous 
line assignments; they may amount to several wavenumbers. But vast 
majority of them amount $0.01 \div 0.1$ cm$^{-1}$ and show random spread 
around "synthesized" values, calculated as differences of optimal 
energy level values. We suppose that they appear due to shifts of 
film blackening maxima for blended lines, finite precision of 
reading from photo plates, and round-up errors in calculating the 
wavenumbers from measured wavelengths in air. 

To refine our knowledge of rovibronic term structure for isotopomers 
of the simplest neutral molecule we launched systematic measurements 
of the rovibronic line wavenumbers in visible part of spectrum 
($400 \div 700$ nm). The $2.65$ m Ebert-Fastie spectrograph was equipped 
with additional camera lens and computer-controlled CMOS matrix 
($22.2$x$14.8$ mm, $1728$x$1152$ pixels) providing the reverse linear 
dispersion about $10^{-3}$ nm/pixel. We used various light sources: 
glow discharge in $Hg$ + $Ar$ (for determination of instrumental 
broadening from the superfine structure of $404.6$ nm and $546.1$ nm 
mercury lines: Gaussian shape with FWHM = $0.183(9)$ and $0.143(9)$
cm$^{-1}$, much smaller than Doppler broadening of $H_2$, $D_2$ lines 
for our gas temperatures $ < 1500$ K), a capillary-arc discharges 
filled with the mixture $D_2$ + $H_2$ + $Ne$ (for studies of the 
precision of spectrometer calibration) 
and with pure $H_2$ or $D_2$ (for final wavenumber measurements). 

We calibrated our spectrometer with available wavelengths in vacuum 
to avoid precise determination of refractive index of air in our 
conditions. The wavenumber values of finite number of strong 
unblended lines of $H_2$ \cite{Dieke1972}, $D_2$ \cite{FSC1985} 
and $Ne$ \cite{Nist} were used as standards. The dependence of 
line wavenumber on the coordinate (in pixels) is monotonic and 
close to linear. The calibration curve of the spectrometer was 
obtained by the least-squares fitting by a second-degree polynomial 
(a linear approximation was insufficient, while a third-degree 
polynomial superfluous). 

The digital intensity registration providing a linear response of 
the detector gave us the opportunity of digital deconvolution of 
the recorded line profiles. For strong unblended lines profiles 
were close to the Gaussian shape except for insignificant far 
wings. Therefore, we approximated all parts of the spectrum 
by Gaussian multipeak fitting with fixed half-width and 
adjustable intensity and wavelength values for peak 
maxima. Thus we obtained wavenumbers of rovibronic 
radiative transitions for triplet-triplet band 
systems.

The contribution contains the details of experimental technique and results.

This work was financially supported in part by the Russian
Foundation for Basic Research, project no. 10-03-00571a.

\end{document}